\documentclass{article}
\usepackage{amsmath,amssymb} 
\usepackage{color}
\usepackage[footnotesize]{caption}
\usepackage{amssymb, epsf, float, graphicx, graphics, verbatim, amssymb,multirow,setspace}

\usepackage{adjustbox}
\usepackage{multirow}
\usepackage{subfigure}

\newcommand{\widthscalefive}{0.145}
\usepackage{multirow}
\usepackage{graphicx}
\usepackage{amsmath,amssymb} 
\usepackage{color}

\title{Image Superresolution using Scale-Recurrent Dense Network} 

\author{Kuldeep Purohit \and
Srimanta Mandal \and
A. N. Rajagopalan}
\date{}
\begin{document}
\maketitle              
\begin{abstract}

Recent advances in the design of convolutional neural network (CNN) have yielded significant improvements in the performance of image super-resolution (SR). The boost in performance can be attributed to the presence of residual or dense connections within the intermediate layers of these networks. The efficient combination of such connections can reduce the number of parameters drastically while maintaining the restoration quality. In this paper, we propose a scale recurrent SR architecture built upon units containing series of dense connections within a residual block (Residual Dense Blocks (RDBs)) that allow extraction of abundant local features from the image. Our scale recurrent design delivers competitive performance for higher scale factors while being parametrically more efficient as compared to current state-of-the-art approaches.
To further improve the performance of our network, we employ multiple residual connections in intermediate layers (referred to as Multi-Residual Dense Blocks), which improves gradient propagation in existing layers.
Recent works have discovered that conventional loss functions can guide a network to produce results which have high PSNRs but are perceptually inferior. We mitigate this issue by utilizing a Generative Adversarial Network (GAN) based framework and deep feature (VGG) losses to train our network. We experimentally demonstrate that different weighted combinations of the VGG loss and the adversarial loss enable our network outputs to traverse along the perception-distortion curve. The proposed networks perform favorably against existing methods, both perceptually and objectively (PSNR-based) with fewer parameters.

\end{abstract}
\section{Introduction}
\label{sec:intro}
Super-resolution (SR) techniques are devised to cope up with the issue of limited-resolution while imaging by generating a high resolution (HR) image from a low resolution (LR) image. However, the possibility of multiple HR images leading to the same LR image makes the problem ill-posed. This can be addressed by regularized mapping of LR image patches to HR counterparts, which are generally extracted from some example images. However, a constrained linear mapping may not be able to represent complex textures of natural images. Deep learning based techniques can behave better in this case by learning a non-linear mapping function.

SR for different scale factors requires separate training of the network. Joint training for
different scale factors can address the issue, as has been attempted by VDSR~\cite{kim2016accurate}, which needs a bicubic interpolated LR image as input. However, this strategy can come in
the way of exploiting hierarchical features from the original LR image, and crucial details may be lost. Further, processing such a high dimensional image for a large number of layers
demands higher computational resources. Another way to deal with the situation is to learn the model for
lower scale factor such as 2 and use it to initialize the learning for higher factors such as 3, 4, etc~\cite{lim2017enhanced}. However, this strategy is parametrically inefficient and does not work well for higher scale factors (e.g., $8$). 

In order to accommodate different up-sampling factors while keeping a check on the number of parameters, we propose a scale-recurrent strategy that helps in transferring learned filters from lower scale factors to higher ones.
We use our scale-recurrent strategy in conjunction with a smaller version of Residual Dense Network (RDN)~\cite{zhang2018residual}, where we use fewer Residual Dense Blocks (RDBs) to reduce the number of parameters as compared to the original RDN. We choose RDBs as building blocks since the combination of residual and dense connections can help in overcoming their individual limitations. This combination allows for efficient flow of information throughout the layers while eliminating the vanishing gradient issue. We refer to this scale-recurrent residual dense network as SRRDN. 

Motivated by the recent developments in network designs based on dense connections, we introduce multiple residual connections within an RDB using $1\times1$ convolutions that results in superior performance with marginal parametric cost. The proposed units are termed as {\it Multi-Residual Dense Blocks (MRDB)}. Our proposed scale-recurrent network with MRDBs is termed as multi-residual dense network (MRDN). 

\section{Related Works}
\label{sec:relate}
Super-resolving a single image generally requires some example HR images to import relevant
information for generating the HR image. Two streams of approaches make
use of the HR example images in their frameworks: i) Conventional, and ii) deep learning based.
The functioning of conventional SR approaches depends on finding patches,
similar to the target patch in the database of patches. Since there could be
many similarities, one needs to regularize the problem. Thus, most of the
conventional approaches focus on discovering regularization techniques in SR such as Tikhinov~\cite{Tikhonov_sr}, total-variation~\cite{TVSR}, Markov random field~\cite{MRF2}, non-local-mean~\cite{Mairal,Glasner,Mandal_SRSI},
sparsity-based prior~\cite{yangj,zeyde2012single,ASDS}, and so on~\cite{NCSR,Mandal_SRSI,rajagopalan2003motion,rajagopalan2005background,suresh2007robust,bhavsar2010resolution,bhavsar2012range,punnappurath2015rolling,punnappurath2017multi,vasu2018analyzing,purohit2020mixed}.

%

Although, the sparsity-based prior works quite efficiently, the linear mapping
of information may fail to represent complex structures of an image. Here, deep-learning based approaches have an upper hand as they can learn a non-linear mapping between LR and corresponding HR image~\cite{dong2016image,dong2016accelerating,wang2015deep,kim2016accurate,tai2017image,lai2017deep,MSLapSRN,sajjadi2017enhancenet,tai2017memnet,lim2017enhanced,zhang2018learning,huang2015single}.
Deep learning stepped into the field of SR via SRCNN~\cite{dong2014learning} by extending the notion of sparse representation using
CNN. The non-linearity involved in CNN is able to better represent complex structures
than conventional approaches to yield superior results. However, going to
deeper architectures increases the difficulty in training such networks. Employing
a residual network into the frame along with skip connections and recursive convolution can mitigate this issue~\cite{kim2016accurate,kim2016deeply}.
Following such an approach, VDSR~\cite{kim2016accurate} and DRCN~\cite{kim2016deeply} methods have demonstrated performance improvement.
The power of recursive blocks involving residual units to create a deeper network was
explored in~\cite{tai2017image}. Recursive unit in conjunction with a gate unit
can act as a memory unit that adaptively combines the previous states with the
current state to produce a super-resolved image~\cite{tai2017memnet}. However, these
approaches interpolate the LR image to the HR grid and feed it to the network.
But this increases the computational requirement due to the higher dimension.

To circumvent the dimension issue, 
networks exists that are tailored to extract features from the LR image which are then processed in subsequent layers. At the end layer, up-sampling is performed to
match with the HR dimension~\cite{dong2016accelerating,ledig2017photo}.
This process can be made faster by reducing the dimension of the features going
to the layers that map from LR to HR and is known as FSRCNN~\cite{dong2016accelerating}. ResNet~\cite{he2016deep} based deeper network
with generative adversarial network (GAN)~\cite{goodfellow2014generative} can produce photo-realistic HR results by including perceptual loss~\cite{johnson2016perceptual} in the network, as devised in SRResNet~\cite{ledig2017photo}. The perceptual loss is further
used with a texture synthesis mechanism in GAN based model to improve SR performance~\cite{sajjadi2017enhancenet}. Though these approaches are able to add textures
in the image, sometimes the results contain artifacts.
The model architecture of SRResNet~\cite{ledig2017photo} has been simplified and
optimized to achieve further improvements in EDSR~\cite{lim2017enhanced}. This was
later modified in MDSR~\cite{lim2017enhanced}, which performs joint training for different scale factors by introducing scale-specific feature extraction and pixel-shuffle layers.

\section{Architecture Design}
\vspace{-3mm}

The success of recent approaches has emphasized the importance of network design. Specifically, most recent image and video SR approaches are built upon two popular image classification networks: residual networks \cite{he2016deep} and densely connected networks \cite{huang2017densely}. These network designs have also enjoyed success and achieved state-of-the-art performance in other image restoration tasks such as image denoising, dehazing, and deblurring. Motivated by the generalization capability of such advances in network designs, the recent work of RDN~\cite{zhang2018residual} proposed a super-resolution network which involves a mixture of both residual and dense connections and yields state-of-the-art results. The fundamental block of this network is RDB, which we too adopt in our work. 

While DenseNet was proposed for high-level computer vision tasks (e.g., object recognition), RDN adopted and improved upon this design to address image SR. Specifically, batch-normalization (BN) layers were removed as they hinder the performance of the network by increasing computational
complexity and memory requirements. The pooling layers are removed too since they could discard important pixel-level information. To enable a higher growth rate, each dense block is terminated with a $1\times1$ conv layer (Local Feature Fusion) and its output is added to the input of the block using Local Residual Learning. This strategy has been demonstrated to be very effective for SR \cite{zhang2018residual}. 

\subsection{Scale-Recurrent Design}
\vspace{-2mm}
Most of the existing SR approaches handle different scale factors independently by neglecting inter-scale relationships. They need to be trained independently for different scale factors. However,
VDSR~\cite{kim2016accurate} can address the issue by jointly training a network for multiple scales. This kind of training requires LR images 
of different resolutions to be up-sampled by bi-cubic interpolation prior
to feeding to the network. Interpolation by a large factor causes loss of
information. Further, it requires higher computational time and memory as
compared to scale-specific networks.

\begin{figure}[!htb]
\vspace{-0.7cm}
\centering
\includegraphics[scale=0.32]{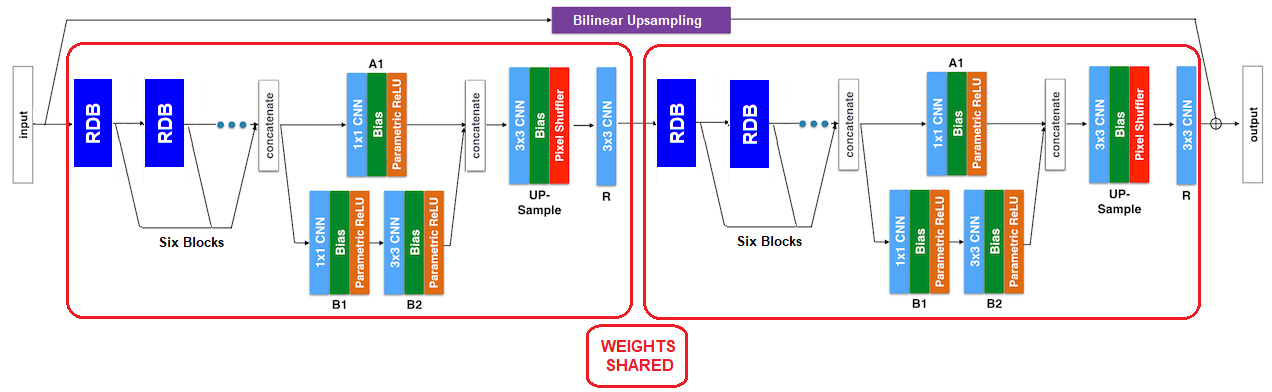}
\vspace{-3mm}
\caption{Network architecture of the proposed Scale-Recurrent Multi-Residual Dense Network for $4\times$ SR.}
\label{fig:network_RCA}
\vspace{-1cm}
\end{figure}

Our network's global design is a multi-scale pyramid which recursively uses the same convolutional filters across scales. This is motivated by the fact that a network capable of super-resolving an image by a factor of 2 can be recursively used to super-resolve the image by a factor $2s, s = 1, 2, 3 \dots$. Even with the same training data, the recurrent exploitation of shared weights works in a way similar to using data multiple times to learn parameters, which actually amounts to data augmentation with respect to scales. We design the network to reconstruct HR images in intermediate steps by progressively performing a $2\times$ upsampling of the input from the previous level. Specifically, we first train a network
to perform SR by a factor of $2$ and then re-utilize the same weights to take the output of $2\times$ as input and result into an output at resolution $4\times$. This architecture is then fine-tuned to perform $4\times$ SR. We experimentally found that such initialization (training for the task of $2\times$ SR) leads to better convergence for larger scale factors. Ours is one of the first approaches to re-utilize the parameters across scales, which significantly reduces the number of trainable parameters while yielding performance gains for higher scale factors.
We term our network SRRDN, whose $4\times$ SR version is shown in Fig. \ref{fig:network_RCA}. \\

\subsection{Multi-Residual Dense Blocks}

We also propose improvements in the structure of RDB for efficient extraction of high-resolution features from low-resolution images. 
The effectiveness of residual and dense connections has been proved in various vision tasks; yet, they cannot be considered as optimum topology.  For example, too many additions on the same feature space may impede  information flow in ResNet~\cite{huang2017densely}. The possibility of  same type of raw features from different layers can lead to redundancies in DenseNet~\cite{chen2017dual}. Some of these issues are addressed in recent image classification networks \cite{chen2017dual,wang2018mixed}. However, these designs are optimized for image classification tasks and their applicability to image restoration has not been explored yet.

\begin{figure}[t]
\centering
\includegraphics[scale=0.22]{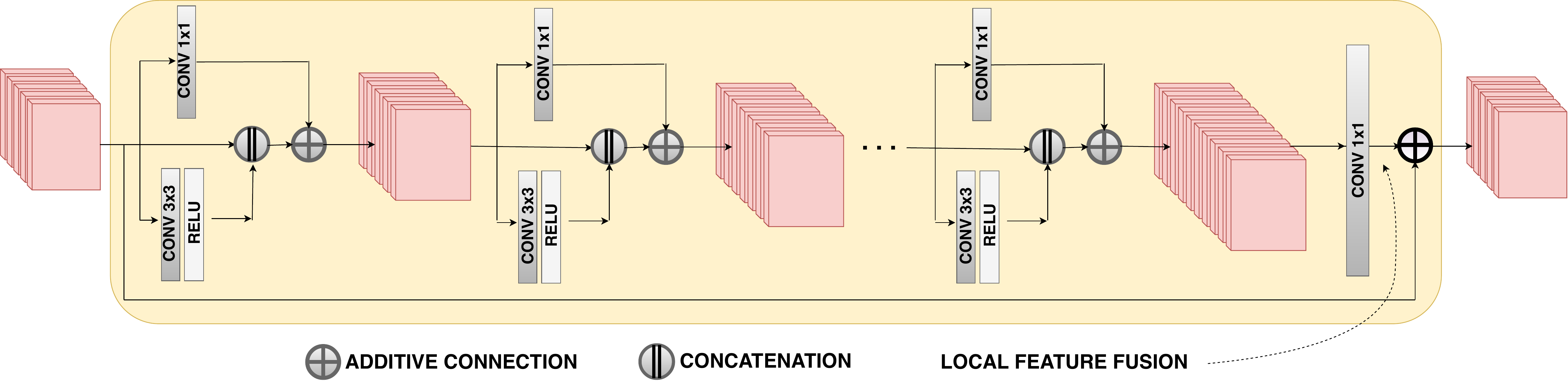}
\vspace{-1mm}
\caption{{\footnotesize Structure of our Multi-Residual Dense Block. Within each module, concatenation operation is performed using the features estimated by the conv $3\times3$ layer, and addition operation is performed on the features estimated by $1\times1$ conv layer (which continuously increase the number of feature maps to match the size of concatenated output). At the end of the block, a $1\times1$ conv layer performs local feature fusion to adaptively control the output information.}}
\label{fig:MRDB}
\vspace{-4mm}
\end{figure}

An RDB of SRRDN already contains multiple paths connecting the current layer to previous network layers. 
One connection is present in the form of a concatenation of features, which is similar to the connections in DenseNet. Although growth rates affect the performance positively, it is harder to train a large number of dense blocks which possess a higher growth rate, as has been experimentally demonstrated in \cite{zhang2018residual}. This can be addressed by Local Feature Fusion (see Fig.~\ref{fig:MRDB}), by including a second connection that stabilizes the training of wide network. This brings down the number of output feature-maps to the number of input feature-maps and enables introduction of a single residual connection between the input and the output of the block (Local Residual Learning).

In order to further improve the gradient flow during training, we introduce a third connection: Multi-Residual connections. Essentially, at each intermediate layer of the block, we convolve the input features using a $1\times1$ conv layer and add them to the output obtained after the concatenation operation.
This type of connection has two properties:  Firstly, existing feature channels get modified, which helps in deeper and hierarchical feature extraction. Secondly, it enables learning of equally meaningful features even with a lower growth-rate during feature concatenation. This strategy promotes new feature exploration with a moderate growth rate and avoids learning of redundant features. These two features enable improved error gradient propagation during training. Our scale-recurrent framework built using MRDBs as basic blocks is termed as MRDN.

\vspace{-5mm}

\section{Experimental Results}
\subsection{Experimental Setup}
\label{subsec:settings}
\vspace{-2mm}
Here, we specify the details of training setup, test data and evaluation metrics. 

\noindent \textbf{Datasets and degradation models.} Following~\cite{timofte2017ntire,lim2017enhanced,zhang2018residual,zhang2018learning}, we use 800 training images from DIV2K dataset~\cite{timofte2017ntire} as training set. For testing, we use five standard benchmark datasets: Set5~\cite{bevilacqua2012low}, Set14~\cite{zeyde2012single}, B100~\cite{martin2001database}, Urban100~\cite{huang2015single}, Manga109~\cite{matsui2017sketch}, and PIRM-self~\cite{2018arXiv180907517B}. We consider bicubic(BI) down-sampling to generate the LR images.

\noindent \textbf{Evaluation metrics.} The SR results are evaluated with two metrics: PSNR and a perceptual metric. For a given image $I$, the perceptual metric is defined as
\begin{equation}
\label{eq:perceptual}
P(I)=\frac{1}{2}((10-M(I)+N(I))
\end{equation}
where $M(I)$ and $N(I)$ are estimated using \cite{ma2017learning} and \cite{mittal2013making}, respectively. These metrics have been used to evaluate different approaches in the PIRM SR Challenge.

\noindent \textbf{Training settings.} Data augmentation is performed on the 800 training images, which are randomly rotated by 90$^{\circ}$, 180$^{\circ}$, 270$^{\circ}$ and flipped horizontally. Our model is trained by ADAM optimizer~\cite{kingma2014adam} with $\beta_{1}=0.9$, $\beta_{2}=0.999$, and $\epsilon=10^{-8}$. The initial leaning rate is set to $10^{-4}$ and is then decreased by half every $2\times10^{5}$ iterations of back-propagation. 

\noindent \textbf{Implementation Details}
The network is implemented using Pytorch library. For training the first network, we used a weighted sum of VGG54 loss and L1 Loss. For the second network, we used a weighted sum of VGG54 loss and conditional-GAN loss. The experiments have been conducted on a machine with i7-4790K CPU, 64GB RAM and 1 NVIDIA Titan X GPU using PyTorch~\cite{paszke2017automatic}. During training, we considered a batch of randomly extracted 16 LR RGB patches of size $32\times32$ pixels. Training the first network (MRDN) took approximately 40 hours. The second network (MRDN-GAN) was then trained for 26 hours.

\subsection{Perceptually Motivated Results}
This work has been used for the purpose of participating in the  PIRM 2018  SR Challenge, which focuses on photo-realistic results (measured using perceptually motivated metric) while maintaining
certain levels of tolerance in terms of root mean squared error (RMSE).
In this challenge, there exist three tracks corresponding to different ranges of RMSE for scale factor of $\times4$. Track 1 corresponds to RMSE $\leq11.5$. Track two: RMSE between $11.5$ and $12.5$, while Track 3 included results with RMSE $\geq12.5$. Perceptually attractive 
images are generally rich in various high-frequency (HF) image details. Thus, the objective is to bring out HF details while super-resolving the given
LR images such that the resultant images yield better perceptual score.
We employed our networks to generate results with scores suitable for each track and proved that our technique can elegantly facilitate quality control during test time. 
Our team \textit{REC-SR} secured the $7^{th}$, $7^{th}$ and  $10^{th}$ ranks in Tracks 1, 2 and 3, respectively.\\\\\\

\vspace{-1cm}
\begin{figure}[!htb]
	\newlength\fsdurthree
	\setlength{\fsdurthree}{-1.5mm}
	\scriptsize
	\centering
	\begin{tabular}{ccc}
		\begin{adjustbox}{valign=t}
		\tiny
			\begin{tabular}{ccc}
				\includegraphics[width=0.309\textwidth]{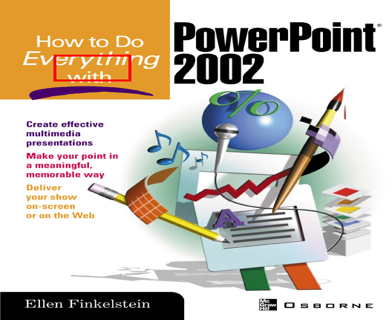}&
				\includegraphics[width=0.309\textwidth]{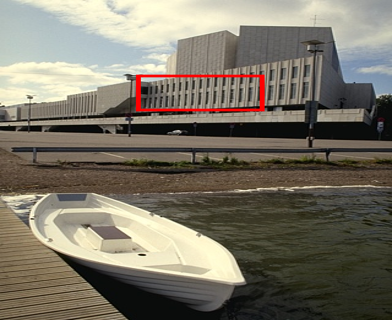}&
				\includegraphics[width=0.309\textwidth]{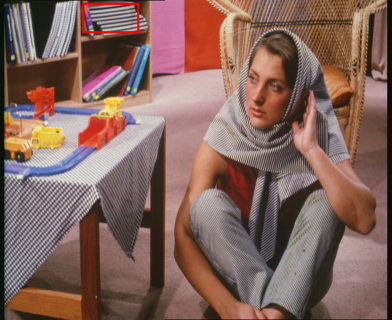}
				
				\\
				Set14 ($4\times$):& BSD100 ($3\times$):& Set14 ($4\times$):\\
				ppt3 & 78004 & Barbara\\			
				
			\end{tabular}
		\end{adjustbox}\\		
		
		\begin{adjustbox}{valign=t}
		\tiny
			\begin{tabular}{cccccc}
				\includegraphics[width=\widthscalefive \textwidth]{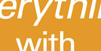} \hspace{\fsdurthree} &
				\includegraphics[width=\widthscalefive \textwidth]{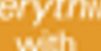} \hspace{\fsdurthree}&
				\includegraphics[width=\widthscalefive \textwidth]{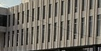} \hspace{\fsdurthree} &
				\includegraphics[width=\widthscalefive \textwidth]{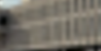} \hspace{\fsdurthree} & 
				\includegraphics[width=\widthscalefive \textwidth]{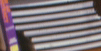} \hspace{\fsdurthree} &
				\includegraphics[width=\widthscalefive \textwidth]{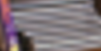} \hspace{\fsdurthree} 
				\\
HR  \hspace{\fsdurthree} & Bicubic \hspace{\fsdurthree} & HR  \hspace{\fsdurthree} & Bicubic  \hspace{\fsdurthree} &HR  \hspace{\fsdurthree} & Bicubic  \hspace{\fsdurthree} 
\\
				\includegraphics[width=\widthscalefive \textwidth]{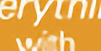} \hspace{\fsdurthree} &
				\includegraphics[width=\widthscalefive \textwidth]{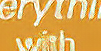} \hspace{\fsdurthree}&
					\includegraphics[width=\widthscalefive \textwidth]{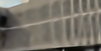} \hspace{\fsdurthree} &
				\includegraphics[width=\widthscalefive \textwidth]{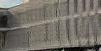} \hspace{\fsdurthree} & 
				\includegraphics[width=\widthscalefive \textwidth]{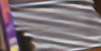} \hspace{\fsdurthree} &
				\includegraphics[width=\widthscalefive \textwidth]{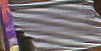} \hspace{\fsdurthree} \\
SRResNet~\cite{ledig2017photo} \hspace{\fsdurthree} &
SRGAN~\cite{ledig2017photo} \hspace{\fsdurthree} &
SRResNet~\cite{ledig2017photo} \hspace{\fsdurthree} &
SRGAN~\cite{ledig2017photo} \hspace{\fsdurthree} &				
SRResNet~\cite{ledig2017photo} \hspace{\fsdurthree} &
SRGAN~\cite{ledig2017photo} \hspace{\fsdurthree} \\				
				
				\includegraphics[width=\widthscalefive \textwidth]{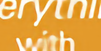} &
						\includegraphics[width=\widthscalefive \textwidth]{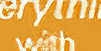} \hspace{\fsdurthree}& 
						\includegraphics[width=\widthscalefive \textwidth]{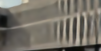} &
						\includegraphics[width=\widthscalefive \textwidth]{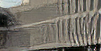} \hspace{\fsdurthree}&
						\includegraphics[width=\widthscalefive \textwidth]{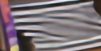} &
						 \includegraphics[width=\widthscalefive \textwidth]{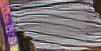} \hspace{\fsdurthree}
						 \\
ENet-E~\cite{sajjadi2017enhancenet}\hspace{\fsdurthree} & ENet-PAT~\cite{sajjadi2017enhancenet} \hspace{\fsdurthree} &ENet-E~\cite{sajjadi2017enhancenet}\hspace{\fsdurthree} & ENet-PAT~\cite{sajjadi2017enhancenet} \hspace{\fsdurthree} &ENet-E~\cite{sajjadi2017enhancenet}\hspace{\fsdurthree} & ENet-PAT~\cite{sajjadi2017enhancenet} \hspace{\fsdurthree}\\
				\includegraphics[width=\widthscalefive \textwidth]{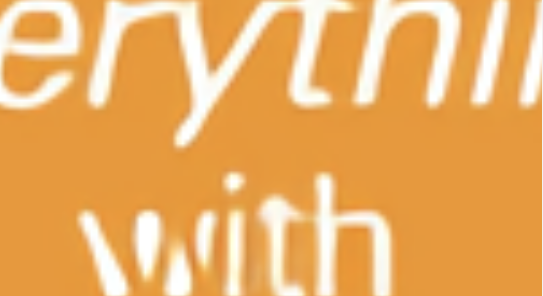} \hspace{\fsdurthree} &
				\includegraphics[width=\widthscalefive \textwidth]{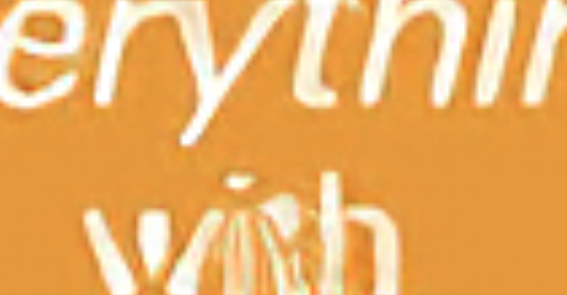} &
				\includegraphics[width=\widthscalefive \textwidth]{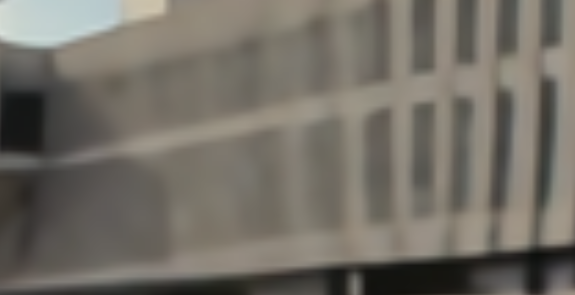} \hspace{\fsdurthree} &
				\includegraphics[width=\widthscalefive \textwidth]{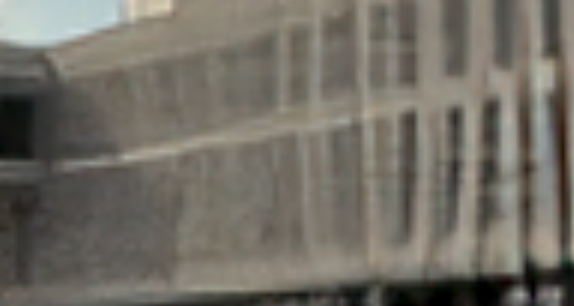} &  
				\includegraphics[width=\widthscalefive \textwidth]{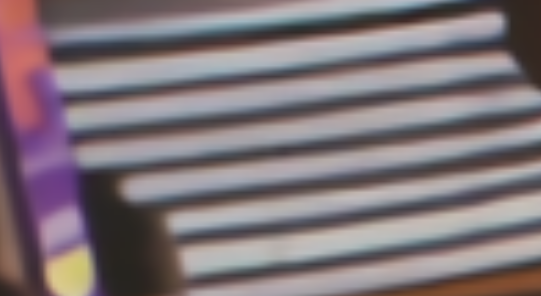} \hspace{\fsdurthree} &
				\includegraphics[width=\widthscalefive \textwidth]{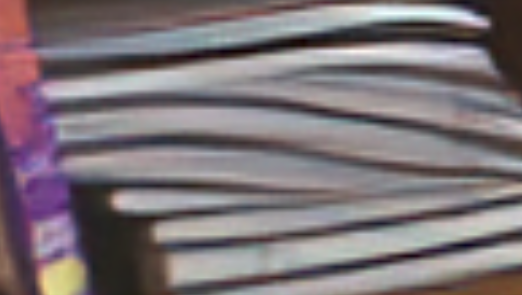}
				\\
MRDN (Ours) \hspace{\fsdurthree} & MRDN-GAN (Ours) \hspace{\fsdurthree}& MRDN (Ours) \hspace{\fsdurthree} & MRDN-GAN (Ours) \hspace{\fsdurthree} & MRDN (Ours) \hspace{\fsdurthree} & MRDN-GAN (Ours)\hspace{\fsdurthree}
\\
			\end{tabular}
		\end{adjustbox}
		\vspace{-3mm}
	\end{tabular}
	\caption{
		Visual comparison for $4\times$ SR on images from Set14 and BSD100 datasets.
	}
\label{fig:result_perceptual}
\vspace{-7mm}
\end{figure}
\begin{figure}[h]
	\newlength\fsdttwofig
	\setlength{\fsdttwofig}{-1.5mm}
	\scriptsize
	\centering
	\begin{tabular}{cc}
		\begin{adjustbox}{valign=t}
		\tiny
			\begin{tabular}{c}
				\includegraphics[width=0.229\textwidth]{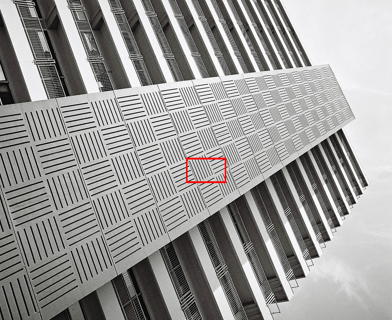}
				\\
				 Urban100 ($4\times$):
				\\
				img\_092
				
			\end{tabular}
		\end{adjustbox}
		\hspace{-2.3mm}
		\begin{adjustbox}{valign=t}
		\tiny
			\begin{tabular}{cccccc}
				\includegraphics[width=\widthscalefive \textwidth]{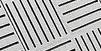} \hspace{\fsdttwofig} &
				\includegraphics[width=\widthscalefive \textwidth]{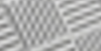} \hspace{\fsdttwofig} &
				\includegraphics[width=\widthscalefive \textwidth]{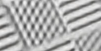} \hspace{\fsdttwofig} &
				\includegraphics[width=\widthscalefive \textwidth]{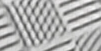} \hspace{\fsdttwofig} &
				\includegraphics[width=\widthscalefive \textwidth]{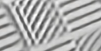}\\
				HR \hspace{\fsdttwofig} &
				Bicubic \hspace{\fsdttwofig} &
				SRCNN~\cite{dong2016image} \hspace{\fsdttwofig} &
				FSRCNN~\cite{dong2016accelerating} \hspace{\fsdttwofig} &
				VDSR~\cite{kim2016accurate} 				
				\\
				\includegraphics[width=\widthscalefive \textwidth]{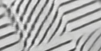} \hspace{\fsdttwofig} &				
				\includegraphics[width=\widthscalefive \textwidth]{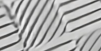} \hspace{\fsdttwofig} &
				\includegraphics[width=\widthscalefive \textwidth]{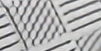} \hspace{\fsdttwofig} &
				\includegraphics[width=\widthscalefive \textwidth]{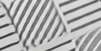} &
				\includegraphics[width=\widthscalefive \textwidth]{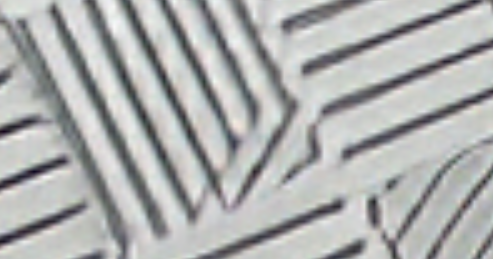}   
				\\ 
				LapSRN~\cite{lai2017deep} \hspace{\fsdttwofig} &
				MemNet~\cite{tai2017memnet} \hspace{\fsdttwofig} &
				SRMDNF~\cite{zhang2018learning} \hspace{\fsdttwofig} &
				EDSR~\cite{lim2017enhanced}  \hspace{\fsdttwofig} &
				MRDN-GAN (Ours)
				\\
			\end{tabular}
		\end{adjustbox}
		\vspace{0.5mm}
		\\
		\begin{adjustbox}{valign=t}
		\tiny
			\begin{tabular}{c}
				\includegraphics[width=0.229\textwidth]{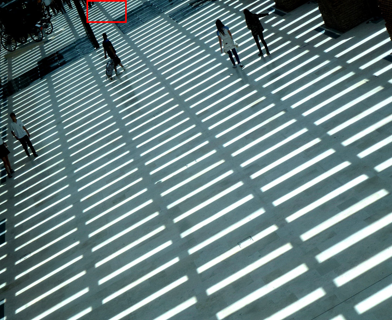}
				\\
				 Urban100 ($4\times$):
				\\
				img\_093
			\end{tabular}
		\end{adjustbox}
		\hspace{-2.3mm}
		\begin{adjustbox}{valign=t}
		\tiny
			\begin{tabular}{cccccc}
				\includegraphics[width=\widthscalefive \textwidth]{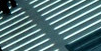} \hspace{\fsdttwofig} &
				\includegraphics[width=\widthscalefive \textwidth]{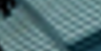} \hspace{\fsdttwofig} &
				\includegraphics[width=\widthscalefive \textwidth]{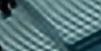} \hspace{\fsdttwofig} &
				\includegraphics[width=\widthscalefive \textwidth]{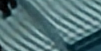} \hspace{\fsdttwofig} &
				\includegraphics[width=\widthscalefive \textwidth]{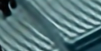} \\
				HR \hspace{\fsdttwofig} &
				Bicubic \hspace{\fsdttwofig} &
				SRCNN~\cite{dong2016image} \hspace{\fsdttwofig} &
				FSRCNN~\cite{dong2016accelerating} \hspace{\fsdttwofig} &
				VDSR~\cite{kim2016accurate}\hspace{\fsdttwofig} &
				\\
				\includegraphics[width=\widthscalefive \textwidth]{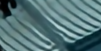} \hspace{\fsdttwofig} &
				\includegraphics[width=\widthscalefive \textwidth]{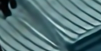} \hspace{\fsdttwofig} &
				\includegraphics[width=\widthscalefive \textwidth]{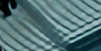} \hspace{\fsdttwofig} &
				\includegraphics[width=\widthscalefive \textwidth]{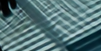} \hspace{\fsdttwofig} &
				\includegraphics[width=\widthscalefive \textwidth]{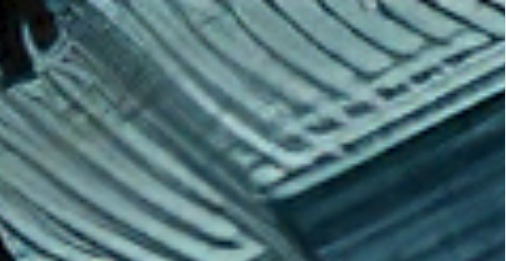}  
				\\ 
				LapSRN~\cite{lai2017deep} \hspace{\fsdttwofig} &
				MemNet~\cite{tai2017memnet} \hspace{\fsdttwofig} &
				SRMDNF~\cite{zhang2018learning} \hspace{\fsdttwofig} &
				EDSR~\cite{lim2017enhanced}  \hspace{\fsdttwofig} &
				MRDN-GAN (Ours)
				\\
			\end{tabular}
		\end{adjustbox}
		\vspace{-4mm}
	\end{tabular}
	\vspace{0.5cm}
	\caption{Visual comparisons for $4\times$ SR on Urban100 dataset.}
\label{fig:result_4x}
\vspace{-8mm}
\end{figure}

\vspace{1cm}
Visual comparisons of the results of our networks MRDN and MRDN-GAN with these techniques on images from standard SR benchmarks are given in Fig.  \ref{fig:result_perceptual}. In all the images, it can be seen that the results of SRResNet and ENetE suffer from blurring artifacts. 
This demonstrates the insufficiency of only pixel-reconstruction losses.
However, the efficient design of our MRDN leads to improved recovery of scene texture in challenging regions. For example, in image ``ppt3", all the compared methods fail to recover the letters `i' and `t'. However, our proposed MRDN recovers them. On the other hand, GAN-based methods of SRGAN, and ENetPAT produce distorted scene textures. The results of ENetPAT are sharper than SRGAN but it generates unwanted artifacts and arbitrary edges (e.g., the result for the image ``78004"). In contrast, our proposed MRDN-GAN leads to textures which are closer to that of the ground-truth HR image too. Similar observations can be found in other images. These comparisons show that the design of SR network plays an important role in both objective and perceptual quality of SR. 

In Fig.~\ref{fig:result_4x}, we compare the results of our model on Urban100 dataset with state-of-the-art SR approaches which are not perceptually motivated for a scale factor of 4. For such texture-rich scenes, a major challenge is to bring out high frequency image details. 
One can observe that most of the existing approaches fail in this aspect and their results are blurred (see Fig. \ref{fig:result_4x}). 
However, our MRDN-GAN is capable of generating sufficiently detailed textures.

\section{Conclusions}

We proposed a scale-recurrent deep architecture, which enables transfer of weights from lower scale factors to the higher ones, in order to reduce the number of parameters as compared to state-of-the-art approaches. We experimentally demonstrated that our scale-recurrent design is well-suited for higher up-sampling factors. The error gradient flow was improved
by elegantly including multiple residual units (MRDN) within the Residual Dense Blocks. To produce perceptually better results, VGG-based loss functions were utilized along with a GAN framework. Different weights were assigned to the loss functions to obtain networks focused on improving either perceptual quality or objective quality during super-resolution. 
The perception-distortion trade-off was addressed by a soft-thresholding technique during test time. We demonstrated the effectiveness of our parametrically efficient model on various datasets.

Refined and complete version of this work appeared in ECCVW 2018.

%
%
%
\bibliographystyle{splncs04}
 \bibliography{egbib}

\end{document}